\documentclass[11pt]{article}
\usepackage{voxturn-report}
\usetikzlibrary{positioning, decorations.pathreplacing, calc}

\newcommand{\vt}{VoxTurn}

\newcommand{\ours}{\rowcolor{vtlight}}
\newcommand{\yes}{\textcolor{vtgood}{$\checkmark$}}
\newcommand{\no}{\textcolor{vtmuted}{--}}

\renewcommand{\reportshorttitle}{\vt{}: several turn-taking questions in one pass}
\title{\vt{}: Answering Several Turn-Taking\\Questions in One Pass with One Model}
\reportsubtitle{An open 0.8B speech LLM for end-of-turn detection and related voice-agent decisions}
\reportauthors{Yang Xiao}
\reportaffiliations{The University of Melbourne}
\reportlinks{%
  \reportlink{\faCube}{Model \& code}{https://huggingface.co/AustinXiao/voxturn}\qquad
  \reportlink{\faPlayCircle}{Interactive demo}{https://swagshaw.github.io/voxturn/}}
\reportdate{October 2026}

\begin{document}

\maketitle

\begin{abstractbox}
\noindent{\sffamily\bfseries\color{vtaccent}Abstract}\par\vspace{3pt}
A cascaded voice agent must decide, throughout each silence of its user, whether the user has finished their turn:
answering too early cuts the user off, answering too late makes the agent slow. It often needs related decisions too:
will the user soon go on, is speech during its own turn an interruption or a backchannel, is a sound speech at all.
With open detectors, each extra decision costs another model or another call, and on the 14-language eot-bench they cut
users off in more than a quarter of mid-turn pauses. Because they are trained on segmented utterances, this mismatch
is the largest single source of error. We present \vt{}, a 0.8B speech LLM that casts these decisions as
multiple-choice questions, scores the hidden state of each option, and answers any set of them in one forward pass
with a block-isolating attention mask, exactly as if asked one at a time. Trained on real conversational pauses with
labels derived from their timing, and re-queried while a silence lasts, \vt{} reaches 11.5\% false cut-offs in
English at a 300\,ms latency budget, comparable to two commercial detectors, and a 14-language mean of 19.9\%, the
lowest among open models, at 8.3\,ms per decision. In Pipecat and LiveKit it cuts off 17\% and 9\% of turns, and it
ranks fifth of 24 in TurnBench end-of-turn detection without any TurnBench training data. The model, code and
framework plugins are released.
\end{abstractbox}

\noindent
\begin{minipage}[t]{0.24\linewidth}\stattile{11.5\%}{English false cut-offs at 300\,ms on eot-bench}\end{minipage}\hfill
\begin{minipage}[t]{0.24\linewidth}\stattile{19.9\%}{14-language mean, lowest among open models}\end{minipage}\hfill
\begin{minipage}[t]{0.24\linewidth}\stattile{8.3\,ms}{per decision; 4 questions in 9.0\,ms}\end{minipage}\hfill
\begin{minipage}[t]{0.24\linewidth}\stattile{\#5\,/\,24}{TurnBench end-of-turn, zero-shot}\end{minipage}

\begin{figure}[p]
  \centering
  \includegraphics[width=\linewidth]{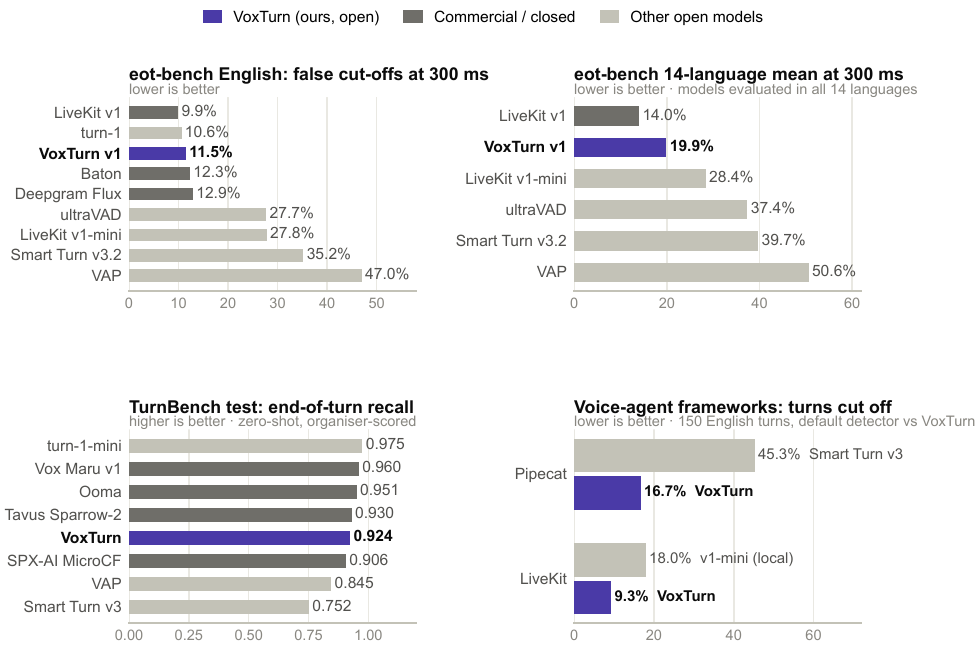}
  \caption{\textbf{\vt{} at a glance.} Top: eot-bench \cite{eotbench}, the share of mid-turn (hold) silences in which
    a detector takes the turn when it must answer within a 300\,ms mean latency budget, in English (left) and averaged
    over 14 languages (right; only systems evaluated in all 14 languages). Bottom left: TurnBench \cite{turnbench}
    test-set end-of-turn recall, scored by the organisers; \vt{} is applied zero-shot. Bottom right: share of 150
    English turns that the agent would cut off, replayed through each framework's VAD and endpointing with its default
    turn detector or with \vt{}. Baseline numbers are official benchmark results or model cards.}
  \label{fig:headline}
  \vspace{0.8em}
  \begin{minipage}{\linewidth}
    \small\sffamily
    \setcounter{tocdepth}{2}
    \renewcommand{\contentsname}{\normalsize Contents}
    \tableofcontents
  \end{minipage}
\end{figure}

\section{Introduction}

Turn-taking is the timing backbone of spoken dialogue \cite{skantze_review}. A cascaded voice agent listens while the
user speaks and, whenever the user falls silent, must decide whether to answer: answering too early cuts the user off
mid-thought, answering too late makes the conversation sluggish. A voice activity detector (VAD) \cite{silero} with a
fixed silence timeout can only trade one error for the other, so recent systems add an end-of-turn (EOT) model that
judges what was said and how. These range from text-only language models over the transcript
\cite{textturn, livekit_td} and small audio classifiers \cite{smartturn} to larger audio models in commercial
products \cite{livekit_td, baton, flux} and in the open \cite{turn1}. EOT-Bench \cite{eotbench} now compares them on
real user turns in 14 languages, and TurnBench \cite{turnbench} on two-party conversations.

Although many works address turn-taking, open detectors leave three problems.
\begin{enumerate}[label=\textbf{\color{vtaccent}\arabic*.}, leftmargin=1.6em]
  \item \textbf{They lag behind commercial detectors, most clearly beyond English.} On eot-bench, their 14-language
    mean is 28\% or more false cut-offs at a 300\,ms budget. We trace much of this gap to training data. Widely used
    training sets consist of segmented utterances, many of them synthetic \cite{smartturn, easyturn}, each labelled
    once at its end, while a deployed detector is asked at varying times inside real pauses, where the answer depends
    on how long the silence has lasted.
  \item \textbf{A deployed agent needs more than one decision}: whether a pause will end soon, whether speech during
    its own turn is an interruption or a backchannel, and whether a sound is speech at all. Each open detector answers
    one of these per call, so every extra answer costs another model or another call.
  \item \textbf{Frameworks query the detector once}, when the VAD reports silence, and fall back to a timeout if the
    answer is not yet confident, although a good model grows more certain as the silence goes on.
\end{enumerate}

\vt{} addresses each problem. For the first, we train on real conversational pauses from phone calls and
face-to-face dialogue, draw each decision time from anywhere in the pause, and derive labels from the pause timing.
For the second, we write each decision as a multiple-choice question for a speech LLM built on Qwen3-ASR-0.6B
\cite{qwen3asr}, score the hidden state of each option instead of generating text, and answer any set of questions in
one forward pass with an attention mask that isolates question blocks after a shared audio prefix, with exactly the
result of asking them one by one. The same pass lets us supervise every training sample with all labels its timing
determines. For the third, we re-query \vt{} every 0.25\,s while the silence lasts.

\begin{highlightbox}{Key results}
\begin{itemize}[leftmargin=1.2em]
  \item \textbf{Real pauses matter most.} Real conversational pauses, with decision times drawn throughout each pause
    and labels derived from its timing, bring most of the gain on eot-bench: English false cut-offs fall from 26.2 to
    16.5\% and the 14-language mean from 33.0 to 20.0\%, more than a speech-aligned backbone, multi-question training
    or four times more training (\S\ref{sec:ablation}).
  \item \textbf{One pass, five decisions.} A one-pass multi-question readout, exactly equivalent to asking the
    questions one by one, lets one 0.8B model answer five turn-taking decisions in 8.3\,ms per decision, and
    four questions in 9.0\,ms instead of 33.2\,ms (\S\ref{sec:mask}, \S\ref{sec:frameworks}).
  \item \textbf{Best open model on eot-bench.} 11.5\% English false cut-offs at 300\,ms, comparable to two commercial
    detectors, and a 19.9\% 14-language mean, the lowest among open models evaluated in all 14 languages
    (\S\ref{sec:eotbench}, \S\ref{sec:multilingual}).
  \item \textbf{Works in real frameworks.} Re-queried during each silence, \vt{} cuts off 17\% and 9\% of turns in
    Pipecat and LiveKit instead of 45\% and 18\%, and answers sooner (\S\ref{sec:frameworks}).
  \item \textbf{Transfers to two-party conversation.} Without TurnBench training, it ranks fifth of 24 in end-of-turn
    and sixth of 23 in interruption detection on the TurnBench test leaderboard (\S\ref{sec:turnbench}).
\end{itemize}
\end{highlightbox}

\section{Related work}

\paragraph{Turn-taking models} Voice Activity Projection \cite{vap} predicts the future voice activity of both
speakers from dialogue audio, and full-duplex speech models such as Moshi \cite{moshi} make turn-taking implicit.
Cascaded agents instead use a dedicated detector on the user channel: TurnGPT \cite{textturn} and the LiveKit text
detector \cite{livekit_td} score the transcript, Smart Turn \cite{smartturn} classifies the last seconds of audio,
and turn-1 \cite{turn1} combines speech-model features with the agent's previous utterance. Easy Turn
\cite{easyturn} and TEN \cite{ten} extend the binary decision to dialogue states such as backchannel and wait, and
the SemanticVAD data \cite{semanticvad} label dialogue text as complete, incomplete, interruption or backchannel.
\vt{} covers these decisions with one model and one pass.

\paragraph{LLMs as classifiers} Scoring answer options instead of generating free text is a standard way to use
language models for multiple-choice questions \cite{lmclassify}. \vt{} applies it to speech, reading a learnt linear
head on the hidden state of each option, and adds a mask that lets several questions share one encoding of the audio.

\paragraph{Benchmarks} eot-bench \cite{eotbench} scores detectors on real single-speaker user turns in 14 languages,
asking for a decision repeatedly as each silence unfolds. TurnBench \cite{turnbench} scores end-of-turn and
interruption events in two-party conversations with one channel per speaker. We report both, together with the test
sets of Easy Turn \cite{easyturn} and TEN \cite{ten} and a nonverbal-vocalisation set \cite{slr99}
(Section~\ref{sec:evalsuite}).

\section{\vt{}}
\label{sec:method}

\begin{figure}[!t]
  \centering
\begin{tikzpicture}[
    font=\sffamily\footnotesize, >=latex, line width=0.5pt,
    tok/.style={draw=white, line width=0.8pt, minimum height=6mm, inner sep=1.5pt, anchor=west, text=vtink},
    box/.style={rounded corners=2pt, minimum height=6mm, inner sep=3pt, align=center},
    lab/.style={font=\sffamily\bfseries\small, text=vtink},
  ]
  \node[tok, fill=vtrule!70, minimum width=7mm]  (hd)  at (0,0) {hdr};
  \node[tok, fill=vtmid!55, minimum width=21mm, right=0pt of hd] (au) {audio ($\times$104)};
  \node[tok, fill=vtrule!70, minimum width=8mm,  right=0pt of au] (tx) {text};
  \node[tok, fill=vtrule!70, minimum width=8mm,  right=0pt of tx] (sp) {spk};
  \node[tok, fill=vtaccent, text=white, minimum width=8mm, right=1.8mm of sp] (q1) {$Q_1$};
  \node[tok, fill=vtaccent!25, minimum width=7mm, right=0pt of q1] (y1) {yes};
  \node[tok, fill=vtaccent!25, minimum width=7mm, right=0pt of y1] (n1) {no};
  \node[tok, fill=vtgray, text=white, minimum width=8mm, right=1.8mm of n1] (q2) {$Q_2$};
  \node[tok, fill=vtgray!30, minimum width=7mm, right=0pt of q2] (o2) {\dots};
  \draw[decorate, decoration={brace, amplitude=4pt, mirror}, vtgray] ([yshift=-2pt]hd.south west) --
      ([yshift=-2pt]sp.south east) node[midway, below=4pt, text=vtgray] {shared prefix (encoded once)};
  \draw[decorate, decoration={brace, amplitude=4pt}, vtgray] ([yshift=2pt]q1.north west) --
      ([yshift=2pt]o2.north east) node[midway, above=4pt, text=vtgray] {question blocks};
  \node[box, fill=vtlight, draw=vtmid, above=6mm of hd.north west, anchor=south west] (enc)
      {8\,s user audio $\rightarrow$ AuT encoder (frozen) $\rightarrow$ projector};
  \draw[->, vtgray] (enc.south -| au.north) -- (au.north);
  \node[box, fill=vtlight, draw=vtaccent, below=9mm of hd.south west, anchor=north west, minimum width=96mm,
        text width=93mm] (llm)
      {\textbf{Qwen3-0.6B + LoRA}: one forward pass with the block-isolating mask};
  \foreach \o in {y1, n1, o2} \draw[->, vtgray] ([yshift=-1pt]\o.south) -- (\o.south |- llm.north);
  \node[box, fill=vtaccent, text=white, below=6mm of llm.south -| y1.east, anchor=north, xshift=-6mm] (head)
      {linear head $w^\top h_{q,o}$, softmax over options $\rightarrow p(\text{yes}),\, p(\text{no})$};
  \draw[->, vtgray] (llm.south -| y1.east) -- (llm.south -| y1.east |- head.north);
  \node[lab, anchor=south west] at ([yshift=2mm]enc.north west) {(a) Sequence and readout};

  \begin{scope}[shift={(114mm,4mm)}]
    \def\s{1.55mm}
    \foreach \i in {0,...,20} {
      \foreach \j in {0,...,20} {
        \pgfmathtruncatemacro{\bi}{\i<10 ? 0 : (\i<16 ? 1 : 2)}
        \pgfmathtruncatemacro{\bj}{\j<10 ? 0 : (\j<16 ? 1 : 2)}
        \pgfmathtruncatemacro{\ok}{(\j<=\i) && ((\bj==0) || (\bi==\bj)) ? 1 : 0}
        \ifnum\ok=1
          \ifnum\bi=0 \def\c{vtmid!60}\else\ifnum\bi=1 \def\c{vtaccent}\else \def\c{vtgray}\fi\fi
          \ifnum\bj=0 \ifnum\bi>0 \def\c{vtmid!60}\fi\fi
          \fill[\c, draw=white, line width=0.3pt] (\j*\s, -\i*\s) rectangle ++(\s, -\s);
        \fi
      }
    }
    \draw[vtrule] (0,0) rectangle (21*\s, -21*\s);
    \foreach \t in {10,16} {
      \draw[vtgray, densely dotted] (\t*\s, 0) -- (\t*\s, -21*\s);
      \draw[vtgray, densely dotted] (0, -\t*\s) -- (21*\s, -\t*\s);
    }
    \node[above, text=vtgray] at (5*\s, 0) {prefix};
    \node[above, text=vtgray] at (13*\s, 0) {$Q_1$};
    \node[above, text=vtgray] at (18.5*\s, 0) {$Q_2$};
    \node[left=1mm, text=vtgray] at (0, -5*\s) {prefix};
    \node[left=1mm, text=vtgray] at (0, -13*\s) {$Q_1$};
    \node[left=1mm, text=vtgray] at (0, -18.5*\s) {$Q_2$};
    \node[below, text=vtgray, font=\sffamily\scriptsize] at (10.5*\s, -21*\s) {keys $\rightarrow$};
    \node[lab, anchor=south west] at (-12mm, 6mm) {(b) Attention mask and position ids};
    \node[anchor=north west, font=\sffamily\scriptsize, inner sep=0pt] at (-12mm, -38mm) {%
      \begin{tabular}{@{}l@{\hspace{4pt}}l@{}}
        \multicolumn{2}{@{}l}{\textbf{position ids}} \\
        prefix & $0, \dots, P{-}1$ \\
        $Q_1$ & $P, \dots, P{+}|Q_1|{-}1$ \\
        $Q_2$ & $P, \dots, P{+}|Q_2|{-}1$ \\
      \end{tabular}};
  \end{scope}
\end{tikzpicture}
  \caption{\textbf{Architecture.} (a) \vt{} reads the last 8\,s of user audio (and optional text) once and answers
    several questions in one forward pass; a linear head scores the hidden state at each option's last token.
    (b) Attention mask (filled: the query row attends to the key column) and position ids: each question block sees
    only the prefix and itself, at the positions it would have if asked alone.}
  \label{fig:arch}
\end{figure}

\subsection{Turn-taking decisions as multiple-choice questions}

At a decision time $t$, \vt{} sees the 8\,s of user audio ending at $t$ (left-padded with silence), optionally a
transcript or the partner's last words, and a marker of who is speaking (user or agent). It answers five questions,
each with a fixed option set (Table~\ref{tab:questions}). Every question has 12 phrasings in English and in Chinese;
training samples one, inference uses a fixed one.

\begin{table}[!h]
  \caption{The five questions \vt{} answers, with their option sets and the training sources that supervise them
    (Table~\ref{tab:data}).}
  \label{tab:questions}
  \centering\small
  \begin{tabularx}{\linewidth}{@{} l >{\hsize=1.25\hsize}X l >{\hsize=0.75\hsize\raggedright\arraybackslash}X @{}}
    \toprule
    \textbf{Question} & \textbf{What it asks} & \textbf{Options} & \textbf{Supervised by} \\
    \midrule
    \texttt{eot}     & Has the user finished their turn?               & yes / no & Smart Turn, pause sources, SemanticVAD \\
    \texttt{resume}  & When will the user go on?                       & \makecell[l]{within 0.5\,s / later /\\ not in this turn} & pause timing (auxiliary) \\
    \texttt{state4}  & Easy Turn dialogue state                        & \makecell[l]{complete / incomplete /\\ backchannel / wait} & Easy Turn \\
    \texttt{intent}  & Is speech during the agent's turn an interruption? & interruption / backchannel & SemanticVAD \\
    \texttt{invalid} & Is the sound not valid input (cough, laughter, noise, music)? & yes / no & MUSAN, VocalSound; real speech as negatives \\
    \bottomrule
  \end{tabularx}
\end{table}

\subsection{Model and readout}

The backbone is Qwen3-ASR-0.6B \cite{qwen3asr}: its AuT audio encoder (frozen) turns 128-bin log-mel features into
13 tokens per second (104 for 8\,s), a projector maps them into Qwen3-0.6B \cite{qwen3}, and we add LoRA
\cite{lora} (rank 16, $\alpha=32$) to every linear layer of the language model. The input sequence is
$[\text{chat header}][\text{audio}][\text{text}][\text{speaker marker}]$ followed by the questions, each written as
``Question: \ldots{} Options: \ldots''. Nothing is generated: a linear head $w$ scores the final hidden state
$h_{q,o}$ of each option $o$ of question $q$, and
\begin{equation}
  p(o \mid q) = \mathrm{softmax}_{o}\big(w^\top h_{q,o}\big)
\end{equation}
over that question's options. Only the projector, the LoRA weights and the head (11.8M parameters) are trained.

\subsection{One-pass multi-question inference}
\label{sec:mask}

Let the prefix occupy positions $1..P$ and question block $k$ occupy a contiguous span $B_k$ after it
(Figure~\ref{fig:arch}b). We use the attention mask
\begin{equation}
  M_{ij} = \mathbf{1}[\, j \le i \,] \cdot \mathbf{1}[\, j \le P \ \lor\ \mathrm{blk}(i) = \mathrm{blk}(j) \,],
\end{equation}
and give the $m$-th token of every block the position id $P+m-1$, the id it would have if its block were the only
question. Prefix tokens never attend to a block, so their states do not depend on the questions; by induction over
layers, every token of $B_k$ then attends to exactly the keys, at exactly the positions, that it would see in the
sequence $[\text{prefix}; B_k]$. With a full-attention backbone, one pass therefore returns the same probabilities
as $K$ separate passes (we measure a maximum difference of $1.5\times10^{-5}$ in fp32), with no dependence on question
order. The prefix ($\approx$120 tokens) is encoded once and each extra question adds about 15 tokens. The argument
does not hold for recurrent or linear-attention layers, whose state cannot be masked per block, which is one reason we
use a full-attention backbone.

\subsection{Training objective}

Each training sample has a main question from its source (Table~\ref{tab:data}). With the one-pass readout we add, at
almost no cost, every auxiliary question whose label is exact: all real speech is labelled \emph{not invalid}, and
every pause sample gets a \texttt{resume} label from its timing. For a decision time $o$ into a pause of length $d$,
a hold pause with remaining silence $d-o \le 0.5$\,s is \emph{within 0.5\,s}, a longer one is \emph{later}, and an
end-of-turn pause is \emph{not in this turn}. We also ask the main question a second time with another phrasing in
the same pass. The loss is
\begin{equation}
  \mathcal{L} = \mathrm{CE}_{\text{main}} + \lambda_{\text{aux}} \overline{\mathrm{CE}}_{\text{aux}}
  + \lambda_{\text{cons}} \tfrac12 \big[\mathrm{KL}(p \,\Vert\, p') + \mathrm{KL}(p' \,\Vert\, p)\big],
\end{equation}
where $p, p'$ are the main question's answers under the two phrasings and $\lambda_{\text{aux}} =
\lambda_{\text{cons}} = 0.5$.

\subsection{Deployment: re-querying during a silence}
\label{sec:deploy}

For serving we merge the LoRA weights, run the encoder on fixed 8\,s windows without host synchronisation, and replay
CUDA graphs captured per input shape. In a voice agent, \vt{} is asked when the VAD reports a pause and then again
every 0.25\,s while the silence lasts; the turn ends as soon as $p(\text{eot}=\text{yes}) \ge \theta$ (default
$\theta = 0.8$), with the framework's timeout as a fallback. This uses the model's growing confidence during a pause,
which a single check discards.

\section{Training data and recipe}
\label{sec:data}

\begin{table}[!t]
  \caption{Training mixture: source, language, main question, size (clips, or pauses for the conversational
    sources) and sampling share. Each pause sample uses a fresh random decision time. Shaded: real conversational
    pauses.}
  \label{tab:data}
  \centering\small
  \setlength{\tabcolsep}{5pt}
  \begin{tabular}{@{} l l l l r r @{}}
    \toprule
    \textbf{Source} & \textbf{Type} & \textbf{Lang.} & \textbf{Question} & \textbf{Size} & \textbf{Share} \\
    \midrule
    Smart Turn v3.2 \cite{smartturn}      & segmented utterances & 23 langs & eot & 266k & 18\% \\
    Easy Turn \cite{easyturn}             & segmented utterances & zh & state4 & 1.19M & 25\% \\
    \ours MagicData-RAMC \cite{ramc}      & conversational pauses & zh & eot & 181k & 11\% \\
    \ours SpokenWOZ \cite{spokenwoz}      & conversational pauses (phone) & en & eot & 95k & 14\% \\
    \ours Seamless Interaction \cite{seamless} & conversational pauses (face-to-face) & en & eot & 179k & 14\% \\
    SemanticVAD \cite{semanticvad}        & dialogue text & en, zh & eot, intent & 15k & 4\% \\
    MUSAN \cite{musan}, VocalSound \cite{vocalsound} & non-speech sounds & -- & invalid & 17k & 14\% \\
    \bottomrule
  \end{tabular}
\end{table}

\paragraph{Data} Table~\ref{tab:data} lists the mixture. Smart Turn and Easy Turn provide segmented utterances, much
of them synthetic. The pause sources are real conversations: for SpokenWOZ (8\,kHz task-oriented phone calls, user
channel) and Seamless Interaction (face-to-face, each speaker's own track) we run Silero VAD \cite{silero} on both
channels and label every silence of at least 100\,ms between two segments of a speaker as \emph{end of turn} if the
other speaker takes the floor ($\ge$1\,s of speech) and as \emph{hold} if nobody speaks before the speaker resumes
within 5\,s; overlaps and backchannels are skipped. RAMC pauses come from its time-aligned transcripts. For every
pause the decision time is drawn uniformly from the first $\min(d, 1.5\,\text{s})$ of the silence, so the model learns
how the answer changes as a pause unfolds. Seamless audio is band-limited to 8\,kHz half of the time so that bandwidth
does not reveal the source. Easy Turn samples are audio only, audio with transcript or transcript only (6:3:1);
invalid negatives are real speech from the other sources; labels within a source are rebalanced by
$n_{\text{label}}^{\,\alpha}$.

\paragraph{Training} We train on 300k samples with batch 128 (4 micro-batches of 32), AdamW (lr $2\times10^{-4}$,
$\beta=(0.9, 0.98)$, weight decay 0.01), 100 warm-up steps and cosine decay to 10\%, gradient clipping at 1.0 and
bf16 autocast (Appendix~\ref{app:hparams}). One run takes 78 minutes on one NVIDIA L40S. Re-running the r3 recipe on
different hardware changed English false cut-offs by 0.1--0.5 points, which we take as the run-to-run noise.

\section{Evaluation suite}
\label{sec:evalsuite}

Table~\ref{tab:evalsuite} summarises what each benchmark measures. The headline benchmark is eot-bench, whose
protocol is worth spelling out because it differs from utterance-level classification.

\begin{table}[!t]
  \caption{Evaluation suite. ``Turns'' are single-speaker user turns; ``FPR'' is the false-positive rate. No
    evaluation audio is used for training.}
  \label{tab:evalsuite}
  \centering\small
  \begin{tabularx}{\linewidth}{@{} l X l X @{}}
    \toprule
    \textbf{Benchmark} & \textbf{What is tested} & \textbf{Languages} & \textbf{Metric} \\
    \midrule
    eot-bench \cite{eotbench} & hold vs.\ end-of-turn silences in real user turns, scored every 100\,ms into each
      silence & 14 & false cut-offs at a 300 / 600\,ms latency budget; latency at 5 / 10\% cut-offs; AUC at
      200\,ms \\
    TurnBench \cite{turnbench} & end-of-turn (EOT) and interruption (INT) events in 30\,h of two-party conversation &
      en & recall at an FPR, median latency \\
    Easy Turn test \cite{easyturn} & four dialogue states from audio & zh & accuracy \\
    TEN test \cite{ten} & finished vs.\ unfinished turns from text & en, zh & per-class recall \\
    OpenSLR 99 \cite{slr99} & nonverbal vocalisations (16 categories) vs.\ 1,786 valid speech clips & -- &
      flagged / falsely rejected \\
    Framework replay & 150 eot-bench English turns through Pipecat \cite{pipecat} and LiveKit Agents
      \cite{livekit_agents} & en & turns cut off, mean response delay \\
    \bottomrule
  \end{tabularx}
\end{table}

\paragraph{The eot-bench protocol} eot-bench \cite{eotbench} contains user turns in 14 languages with their silences
labelled \emph{hold} (the user goes on) or \emph{end of turn}. A model is scored every 100\,ms into each silence
(minimum silence 100\,ms), and the eot-harness sweeps decision policies (threshold, action delay, timeout). A
\emph{false cut-off} is a hold silence in which the policy takes the turn; the \emph{latency} is the mean time from
the start of an end-of-turn silence to the response. For each detector we report the lowest false cut-off rate at a
mean response latency of 300\,ms and 600\,ms, the lowest mean latency at 5\% and 10\% cut-offs, and the AUC between
hold and end-of-turn silences 200\,ms into the pause. Every number is therefore a point on a trade-off curve
(Figure~\ref{fig:curve}a), and lower is better for both cut-offs and latency.

\paragraph{Baselines} Baseline numbers are the official eot-bench results; turn-1's are from its model card
\cite{turn1}. TurnBench numbers are from the organisers' test leaderboard (October 2026). eot-bench audio is never
used for training.

\FloatBarrier
\section{Results}

\subsection{eot-bench English: the latency--accuracy trade-off}
\label{sec:eotbench}

\begin{table}[!t]
  \caption{\textbf{eot-bench.} English false cut-offs (\%) at a mean-latency budget and mean latency (ms) at a
    cut-off budget; Chinese and the 14-language mean at 300\,ms. Lower is better everywhere. --: not supported or
    not reported. Sorted by English @300.}
  \label{tab:main}
  \centering\small
  \setlength{\tabcolsep}{5pt}
  \begin{tabular}{@{} l c c c c c c c c @{}}
    \toprule
    & & & \multicolumn{4}{c}{\textbf{English}} & \textbf{Chinese} & \textbf{14 lang.} \\
    \cmidrule(lr){4-7} \cmidrule(lr){8-8} \cmidrule(l){9-9}
    & & & \multicolumn{2}{c}{cut-offs (\%) at} & \multicolumn{2}{c}{latency (ms) at} & & \\
    \textbf{System} & \textbf{Open} & \textbf{Params} & \textbf{300\,ms} & \textbf{600\,ms} & \textbf{5\%} &
      \textbf{10\%} & \textbf{@300} & \textbf{mean} \\
    \midrule
    LiveKit v1 \cite{livekit_td}        & \no  & --   & 9.9  & 4.5  & 543  & 295  & 20.6 & 14.0 \\
    turn-1 \cite{turn1}                 & \yes & 2.8B & 10.6 & 4.5  & 553  & 322  & --   & -- \\
    \ours \textbf{\vt{} v1}             & \yes & 0.8B & \textbf{11.5} & \textbf{6.0} & \textbf{669} & \textbf{335} & \textbf{22.9} & \textbf{19.9} \\
    Baton \cite{baton}                  & \no  & --   & 12.3 & 4.8  & 577  & 350  & --   & -- \\
    Deepgram Flux \cite{flux}           & \no  & --   & 12.9 & 9.9  & 1151 & 548  & --   & -- \\
    ultraVAD \cite{ultravad}            & \yes & --   & 27.7 & 11.9 & 899  & 663  & 38.8 & 37.4 \\
    LiveKit v1-mini \cite{livekit_td}   & \yes & --   & 27.8 & 12.1 & 1071 & 698  & 27.5 & 28.4 \\
    Smart Turn v3.2 \cite{smartturn}    & \yes & --   & 35.2 & 14.8 & 1051 & 739  & 42.9 & 39.7 \\
    VAP \cite{vap}                      & \yes & --   & 47.0 & 14.6 & 1131 & 749  & 51.2 & 50.6 \\
    \bottomrule
  \end{tabular}
\end{table}

\begin{figure}[!t]
  \centering
  \includegraphics[width=\linewidth]{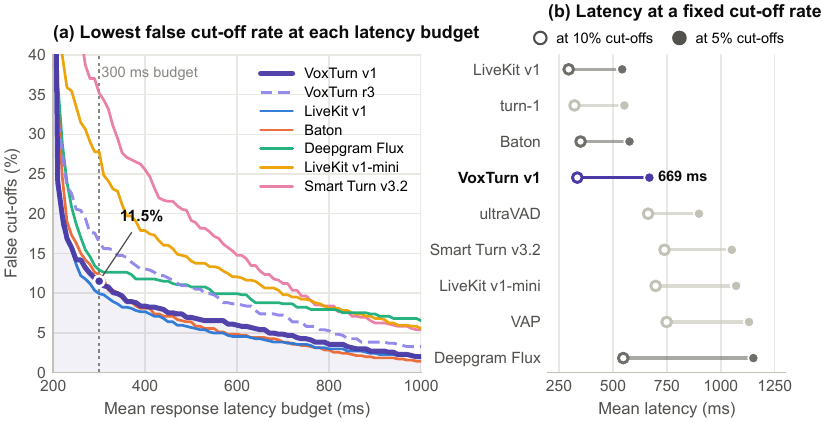}
  \caption{\textbf{eot-bench English.} (a) Lowest false cut-off rate at each mean-latency budget, from the
    eot-harness policy sweep; the shaded area is under the \vt{} v1 curve. \vt{} r3 is our earlier Whisper-based
    model (Table~\ref{tab:ablation}). (b) Mean latency each system needs to keep false cut-offs at 10\% (open
    circle) and 5\% (filled circle); shorter is better.}
  \label{fig:curve}
\end{figure}

Table~\ref{tab:main} compares \vt{} with the systems on eot-bench. In English, \vt{} cuts the user off in
\textbf{11.5\%} of hold silences at a 300\,ms latency budget, fewer than the commercial Baton (12.3\%) and Deepgram
Flux (12.9\%) and behind only LiveKit's cloud detector and the 3.5$\times$ larger turn-1.

Figure~\ref{fig:curve}a shows the full trade-off. Three groups emerge. LiveKit v1, Baton and \vt{} form the
lowest band at every budget; Deepgram Flux is close to them at tight budgets but flattens beyond about 500\,ms; and the
other open models (LiveKit v1-mini, Smart Turn v3.2) need 1.6--2.2$\times$ the latency of \vt{} for the same
5\% or 10\% cut-off rate. At
tight budgets \vt{} tracks the commercial detectors, and it stays far below the other open models at every budget.
Figure~\ref{fig:curve}b reads the same curves the other way: at 5\% cut-offs \vt{} answers after 669\,ms, about
400\,ms sooner than Smart Turn and LiveKit v1-mini, and at 10\% cut-offs after 335\,ms, within 40\,ms of LiveKit
v1.

\subsection{Multilingual results}
\label{sec:multilingual}

\begin{figure}[!t]
  \centering
  \includegraphics[width=\linewidth]{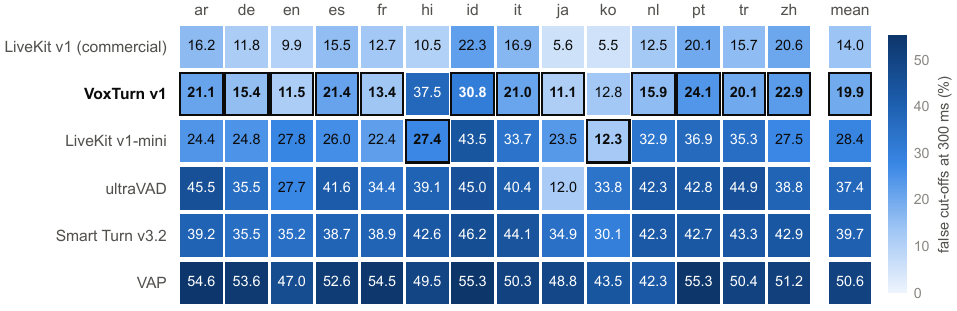}
  \caption{\textbf{eot-bench, 14 languages}: false cut-offs (\%) at a 300\,ms latency budget for every system
    evaluated in all 14 languages, rows sorted by the mean. Outlined: the best open model in each column. Only English
    and Chinese (purple) have conversational-pause training data.}
  \label{fig:multilingual}
\end{figure}

Across the 14 languages \vt{} has a mean of \textbf{19.9\%}, against 28.4\% for the next open model
(Table~\ref{tab:main}, Figure~\ref{fig:multilingual}). It is the best open model in 12 of the 14 languages; only in
Hindi and Korean is LiveKit v1-mini lower (27.4 vs.\ 37.5\% and 12.3 vs.\ 12.8\%). Its strongest languages are
Japanese (11.1\%), English (11.5\%), Korean (12.8\%) and French (13.4\%), and ten of the 14 languages are below
22\%. Read at a fixed 5\% cut-off rate instead, \vt{} answers sooner than every other open model in all 14
languages (Appendix~\ref{app:lang}, Table~\ref{tab:lang}).

The multilingual gain came with the real pause data: adding English and Chinese conversational pauses (r0 $\to$ r3
in Table~\ref{tab:ablation}) lowered the 14-language mean from 33.0\% to 20.0\%, although no pause data was added in
the other twelve languages. The pause cues learnt from these conversations therefore transfer to other languages.

\subsection{What drives the gains}
\label{sec:ablation}

\begin{table}[!t]
  \caption{\textbf{The path to v1} (eot-bench, 300k training samples each). Rows r0--r3 use Whisper-small
    \cite{whisper} with Qwen3.5-0.8B; r3$^\dagger$ is r3 re-run on the L40S, the control for the rows below it.}
  \label{tab:ablation}
  \centering\small
  \setlength{\tabcolsep}{5pt}
  \begin{tabular}{@{} l c c c c c @{}}
    \toprule
    & \multicolumn{3}{c}{\textbf{English}} & \textbf{Chinese} & \textbf{14 lang.} \\
    \cmidrule(lr){2-4} \cmidrule(lr){5-5} \cmidrule(l){6-6}
    \textbf{Configuration} & \textbf{@300 (\%)} $\downarrow$ & \textbf{@5\% (ms)} $\downarrow$ & \textbf{AUC} $\uparrow$ &
      \textbf{@300 (\%)} $\downarrow$ & \textbf{mean (\%)} $\downarrow$ \\
    \midrule
    r0: segmented utterances, RAMC at 0.2\,s  & 26.2 & 921 & .896 & 42.1 & 33.0 \\
    r1: + RAMC decision time within pause     & 22.6 & 936 & .901 & 32.7 & -- \\
    r2: + SpokenWOZ pauses                    & 16.3 & 850 & .944 & 29.1 & -- \\
    r3: + Seamless pauses                     & 16.5 & 808 & .940 & 26.5 & 20.0 \\
    \quad 4$\times$ more training (1.2M)      & 15.9 & 833 & .948 & 24.8 & 21.0 \\
    \midrule
    r3$^\dagger$                              & 16.6 & 855 & .939 & 26.3 & 20.5 \\
    \quad MLP head + pitch branch             & 18.4 & 966 & .927 & 29.6 & 21.6 \\
    \quad Qwen3-ASR-0.6B backbone             & 12.8 & 709 & .961 & 21.9 & 19.7 \\
    \ours \quad + multi-question (\textbf{v1}) & \textbf{11.5} & \textbf{669} & \textbf{.963} & 22.9 & 19.9 \\
    \bottomrule
  \end{tabular}
\end{table}

\begin{figure}[!t]
  \centering
  \includegraphics[width=\linewidth]{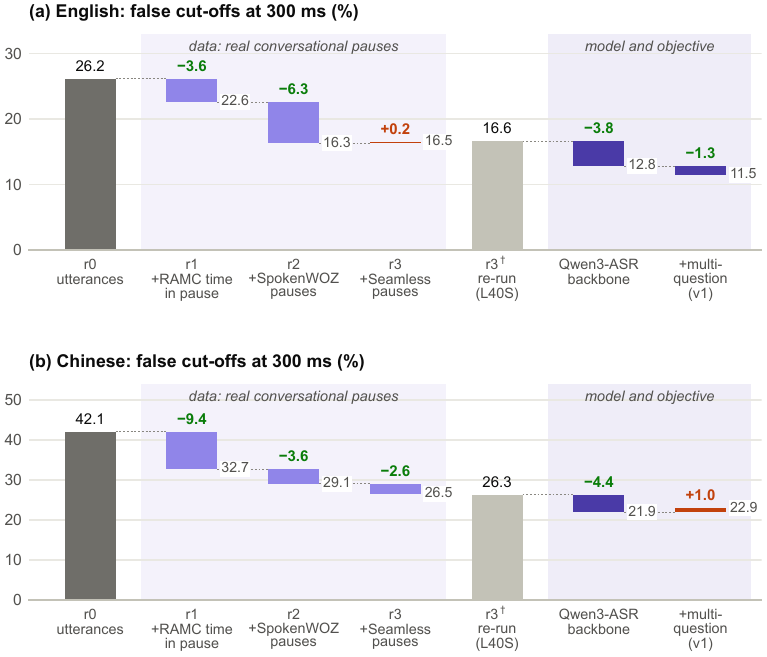}
  \caption{\textbf{The path to v1 as a waterfall.} Floating bars show the change in false cut-offs at 300\,ms from
    the step before; the number above each gives the change (green: fewer cut-offs, orange: more) and the number
    beside it the new value. Grey bars are absolute values. r3$^\dagger$ re-runs r3 on
    different hardware and restarts the chain. Data steps (shaded left) bring the largest gains in both languages.}
  \label{fig:ablation}
\end{figure}

Table~\ref{tab:ablation} and Figure~\ref{fig:ablation} trace the development.

\paragraph{Data first} With segmented utterances (r0), AUC stayed below 0.90 whatever we tuned. Drawing RAMC
decision times from anywhere in a pause (r1) helped Chinese most ($-$9.4 points), and adding real English pauses from
task-oriented phone calls (r2) gave the largest single English gain: 26.2 $\to$ 16.3\% false cut-offs and AUC 0.896
$\to$ 0.944. Face-to-face pauses (r3) added little in English but improved Chinese and latency. The pause data, not
the amount of training, is what matters: four times more updates gain only 0.7 points, and neither a larger readout
head nor an explicit pitch branch helps.

\paragraph{Then the backbone} The second large step is the backbone. Replacing Whisper-small and a lightly trained
adapter with Qwen3-ASR-0.6B, whose encoder and language model are already aligned on large-scale speech, lowers
English cut-offs from 16.6 to 12.8\% and the Chinese calibration error of $p(\text{eot})$ on eot-bench from 0.30 to
0.11.

\paragraph{Then multi-question training} Multi-question training improves English at every latency budget between
250 and 500\,ms by 0.4--1.8 points (11.5\%, AUC 0.963) while adding the resume, invalid and consistency supervision;
the English $p(\text{eot})$ is well calibrated (ECE 0.007). In Chinese it costs 1.0 point at 300\,ms and the
14-language mean is nearly unchanged (19.7 vs.\ 19.9\%).

\subsection{Other decisions from the same pass}
\label{sec:other}

\begin{figure}[!t]
  \centering
  \includegraphics[width=\linewidth]{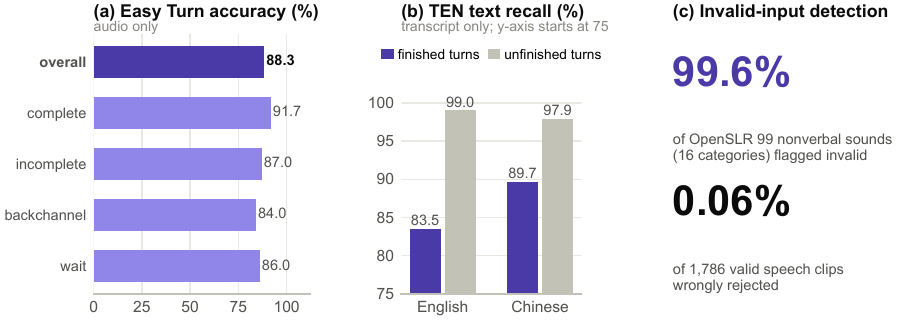}
  \caption{\textbf{Other questions answered by the same model and pass.} (a) Easy Turn test set, four dialogue
    states from audio alone. (b) TEN text test set, recall of finished and unfinished turns. (c) Invalid-input
    detection on OpenSLR 99 nonverbal vocalisations and on valid speech.}
  \label{fig:other}
\end{figure}

The same model and pass also answer the other questions (Figure~\ref{fig:other}). On the Easy Turn test set, \vt{}
reaches 88.3\% accuracy over the four dialogue states (complete 91.7, incomplete 87.0, backchannel 84.0, wait 86.0)
from audio alone. It flags 99.6\% of the OpenSLR 99 nonverbal vocalisations (coughs, laughter, sighs and 13 other
categories) as invalid input while rejecting only 0.06\% of 1,786 valid speech clips. On TEN's text test set it
recognises 83.5\% (English) and 89.7\% (Chinese) of finished turns and 99.0\% and 97.9\% of unfinished ones.

\subsection{Two-party conversations: TurnBench}
\label{sec:turnbench}

\begin{table}[!t]
  \caption{\textbf{TurnBench.} Recall (R) at a false-positive rate (FPR) and median latency (ms). Top: test set,
    scored by the organisers (October 2026; $\dagger$: external submission); ranks are among all submissions.
    Bottom: dev (38 conversations, FPR $\le 0.1$) and EOT policy ablations re-searched on dev.}
  \label{tab:turnbench}
  \centering\small
  \setlength{\tabcolsep}{6pt}
  \begin{tabular}{@{} l c c r c c r @{}}
    \toprule
    & \multicolumn{3}{c}{\textbf{EOT}} & \multicolumn{3}{c}{\textbf{INT}} \\
    \cmidrule(lr){2-4} \cmidrule(l){5-7}
    \textbf{System} & \textbf{R} $\uparrow$ & \textbf{FPR} & \textbf{ms} $\downarrow$ & \textbf{R} $\uparrow$ &
      \textbf{FPR} & \textbf{ms} $\downarrow$ \\
    \midrule
    turn-1-mini$^\dagger$ \cite{turn1mini} & .975 & .075 & 595  & .973 & .048 & 551 \\
    Vox Maru v1$^\dagger$              & .960 & .072 & 548  & .964 & .085 & 368 \\
    Ooma$^\dagger$                     & .951 & .082 & 388  & .968 & .076 & 567 \\
    Tavus Sparrow-2$^\dagger$          & .930 & .084 & 615  & .958 & .126 & 959 \\
    SPX-AI MicroCF$^\dagger$           & .906 & .079 & 678  & .984 & .073 & 747 \\
    VAP \cite{vap}                     & .845 & .055 & 368  & .945 & .107 & 994 \\
    Smart Turn v3 \cite{smartturn}     & .752 & .047 & 1017 & .107 & .093 & 159 \\
    \ours \textbf{\vt{}} (EOT \#5 of 24, INT \#6 of 23) & \textbf{.924} & .076 & 557 & \textbf{.951} & .112 & 332 \\
    \midrule
    \multicolumn{7}{@{}l}{\textit{Dev set, EOT}} \\
    \vt{}, threshold, own channel      & .700 & .100 & 597  & .960 & .100 & 337 \\
    \quad two-channel score            & .810 & .100 & 351  & & & \\
    \quad + decision policy            & .930 & .094 & 565  & & & \\
    Policy, re-searched                & .925 & .087 & 667  & & & \\
    \quad $-$ confirmation event       & .895 & .087 & 762  & & & \\
    \quad $-$ partner trigger          & .885 & .098 & 751  & & & \\
    \quad VAD in place of \vt{}        & .924 & .098 & 821  & & & \\
    \bottomrule
  \end{tabular}
\end{table}

\begin{figure}[!t]
  \centering
  \includegraphics[width=\linewidth]{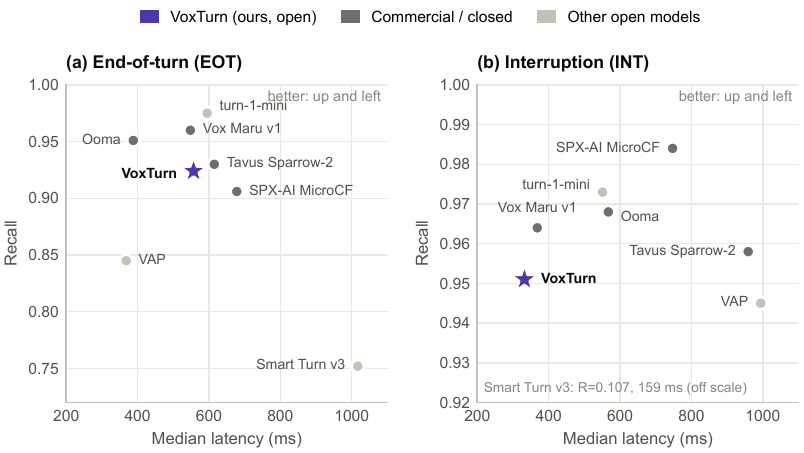}
  \caption{\textbf{TurnBench test set}: recall against median latency (upper left is better). Systems operate at
    different false-positive rates (Table~\ref{tab:turnbench}). Smart Turn v3's INT recall (0.107) is off the scale
    of panel (b).}
  \label{fig:turnbench}
\end{figure}

TurnBench \cite{turnbench} has 30\,h of triple-annotated two-party English conversations, one channel per speaker.
Systems emit causal event times for \textbf{EOT} (a turn has ended) and \textbf{INT} (speech over the partner is a
floor-taking interruption, not a backchannel); an event counts within $[-0.25, +3]$\,s of a gold event, and firing in
a mid-turn pause or backchannel is one false positive per span. \vt{} runs zero-shot: at 10\,Hz per channel it answers
\texttt{eot} and \texttt{resume} in silences, and \texttt{state4} and \texttt{invalid} in the first 2.5\,s of speech
begun near the partner's; INT thresholds $p(\text{not backchannel})\,p(\text{not invalid})$. For EOT we add a
two-channel decision policy, described in Appendix~\ref{app:policy}, whose ten parameters are random-searched for the
best dev recall at FPR $\le 0.1$.

\paragraph{Test leaderboard} On the held-out test set (Table~\ref{tab:turnbench}, Figure~\ref{fig:turnbench}), \vt{}
ranks fifth of 24 in EOT (0.924) and sixth of 23 in INT (0.951), and is the fastest of the INT top six (332\,ms),
well ahead of the strongest baseline in \cite{turnbench}, VAP. turn-1-mini, first in EOT, reports 22.6\% English false
cut-offs on eot-bench \cite{turn1mini} against 11.5\% for \vt{}.

\begin{figure}[!t]
  \centering
  \includegraphics[width=\linewidth]{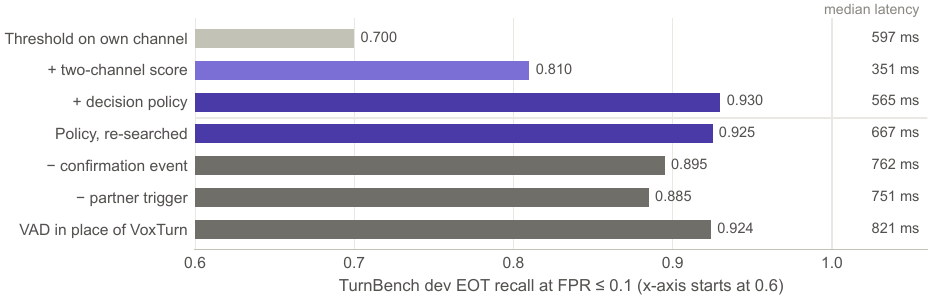}
  \caption{\textbf{TurnBench dev, EOT decision policy.} Recall at FPR $\le 0.1$ as the policy is built up (top three)
    and when its parts are removed from the re-searched policy (bottom three); the right column gives the median
    latency.}
  \label{fig:policy}
\end{figure}

\paragraph{What the policy adds} On dev (Figure~\ref{fig:policy}), the partner's speech lifts the own-channel EOT
recall from 0.700 to 0.810, and the full policy to 0.930; the partner trigger and the confirmation event add 4.0 and
3.0 points. With VAD in place of \vt{} the recall holds (0.924) but answers come later (821 vs.\ 667\,ms).

\subsection{Efficiency and use in voice-agent frameworks}
\label{sec:frameworks}

\begin{table}[!t]
  \caption{\textbf{Efficiency and framework replay.} Left: latency per call (L40S, bf16, 8\,s audio, end to end,
    p50). Right: 150 eot-bench English turns through each framework's Silero VAD and endpointing (LiveKit: VAD minimum
    silence 0.3\,s); \vt{} re-checks every 0.25\,s. A turn is cut off if any of its pauses ends it; delay is the mean
    from the final silence.}
  \label{tab:deploy}
  \centering\small
  \begin{minipage}[t]{0.44\linewidth}\centering
  \begin{tabular}[t]{@{} l r @{}}
    \toprule
    \textbf{Call} & \textbf{ms} \\
    \midrule
    1 question, unmerged eager model & $\approx$38 \\
    \ours 1 question, serving mode & \textbf{8.3} \\
    4 questions in 4 passes & 33.2 \\
    \ours 4 questions in one pass & \textbf{9.0} \\
    4 clips in one batch, 1 question each & 16.0 \\
    \bottomrule
  \end{tabular}
  \end{minipage}\hfill
  \begin{minipage}[t]{0.53\linewidth}\centering
  \begin{tabular}[t]{@{} l l r r @{}}
    \toprule
    \textbf{Framework} & \textbf{Turn detector} & \textbf{Cut off} & \textbf{Delay} \\
    \midrule
    \multirow{2}{*}{Pipecat}  & Smart Turn v3 (default) & 45.3\% & 1.32\,s \\
                              & \cellcolor{vtlight}\vt{} & \cellcolor{vtlight}\textbf{16.7\%} & \cellcolor{vtlight}\textbf{0.36\,s} \\
    \midrule
    \multirow{2}{*}{LiveKit}  & v1-mini (local)         & 18.0\% & 0.73\,s \\
                              & \cellcolor{vtlight}\vt{} & \cellcolor{vtlight}\textbf{9.3\%} & \cellcolor{vtlight}\textbf{0.47\,s} \\
    \bottomrule
  \end{tabular}
  \end{minipage}
\end{table}

\begin{figure}[!t]
  \centering
  \includegraphics[width=\linewidth]{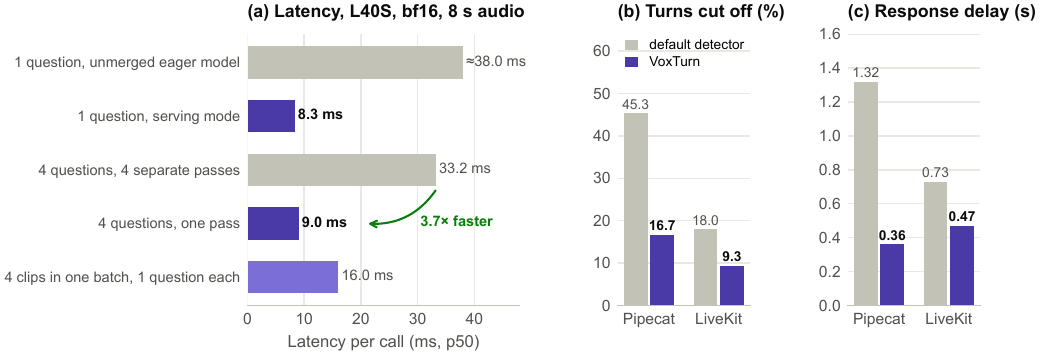}
  \caption{\textbf{Efficiency and deployment.} (a) Latency per call; answering four questions in one pass costs about
    as much as one question. (b, c) Framework replay of 150 English eot-bench turns: share of turns the agent would
    cut off and mean delay from the start of the final silence, with each framework's default local detector and with
    \vt{} re-checked every 0.25\,s.}
  \label{fig:efficiency}
\end{figure}

\paragraph{Latency} One decision takes \textbf{8.3\,ms} in serving mode (Table~\ref{tab:deploy}, 1.9\,GB of GPU
memory), and four questions in one pass take 9.0\,ms instead of 33.2\,ms, a 3.7$\times$ saving whose answers differ
from separate passes only by bf16 rounding. Serving mode keeps the eot-bench result (11.6\% vs.\ 11.5\% at 300\,ms).

\paragraph{Framework replay} We integrated \vt{} into Pipecat \cite{pipecat} as a turn analyzer and into LiveKit
Agents \cite{livekit_agents} as a streaming audio turn detector, both with the 0.25\,s re-check policy, and replayed
150 English eot-bench turns through each framework's VAD and endpointing rules. A turn counts as cut off if the agent
would have answered in any of its mid-turn pauses; delay is measured from the start of the final silence. In Pipecat,
\vt{} cuts off \textbf{16.7\%} of turns against 45.3\% with the default Smart Turn v3 and answers after 0.36\,s
instead of 1.32\,s on average. In LiveKit, with the VAD's minimum silence lowered to 0.3\,s, it cuts off half as many
turns as LiveKit's local audio detector (\textbf{9.3\%} vs.\ 18.0\%) and answers sooner (0.47\,s vs.\ 0.73\,s; 90th
percentile 0.78\,s vs.\ 2.50\,s).

\paragraph{The re-check matters} Asked only once (threshold 0.5), \vt{} in Pipecat waits 1.26\,s on average, because
a pause that is not yet convincing when the VAD reports it falls back to the framework's 3\,s timeout. Replays of such
turns in six languages, with both models' probabilities over time, can be explored in the interactive
demo.\footnote{\url{https://swagshaw.github.io/voxturn/}}

\FloatBarrier
\section{Conclusion}
\label{sec:conclusion}

\vt{} answers several turn-taking questions with a speech LLM in one forward pass. Real conversational pauses, a
speech-aligned backbone and supervision of every label a pause's timing determines bring an open 0.8B model to 11.5\%
English false cut-offs at 300\,ms and the best 14-language mean among open models, at 8.3\,ms per decision, and to
fifth place in TurnBench end-of-turn detection. Weights, code and plugins are at
\url{https://huggingface.co/AustinXiao/voxturn}, and the demo at \url{https://swagshaw.github.io/voxturn/}.

\section*{Acknowledgements}

Experiments used computational resources of the National Computational Infrastructure (NCI Australia) and of the
University of Melbourne. Generative AI tools were used to assist with writing code, running and summarising
experiments, and drafting and editing this report; the authors reviewed and take responsibility for all content.

{\small
\bibliographystyle{unsrtnat}
\bibliography{refs}
}

\clearpage
\appendix
\section{Per-language eot-bench results}
\label{app:lang}

\begin{table}[!h]
  \caption{eot-bench per language: false cut-offs (\%) at a 300\,ms mean-latency budget (top) and mean latency (ms) at 5\% false cut-offs (bottom); lower is better. Bold: best open model per column.}
  \label{tab:lang}
  \centering\footnotesize
  \setlength{\tabcolsep}{3.1pt}
  \begin{tabular}{@{} l rrrrrrrrrrrrrr r @{}}
    \toprule
    & \textbf{ar} & \textbf{de} & \textbf{en} & \textbf{es} & \textbf{fr} & \textbf{hi} & \textbf{id} & \textbf{it} & \textbf{ja} & \textbf{ko} & \textbf{nl} & \textbf{pt} & \textbf{tr} & \textbf{zh} & \textbf{mean} \\
    \midrule
    \multicolumn{16}{@{}l}{\textit{False cut-offs at 300 ms (\%)}} \\
    LiveKit v1 (closed) & 16.2 & 11.8 & 9.9 & 15.5 & 12.7 & 10.5 & 22.3 & 16.9 & 5.6 & 5.5 & 12.5 & 20.1 & 15.7 & 20.6 & 14.0 \\
    \ours \textbf{VoxTurn v1} & \textbf{21.1} & \textbf{15.4} & \textbf{11.5} & \textbf{21.4} & \textbf{13.4} & 37.5 & \textbf{30.8} & \textbf{21.0} & \textbf{11.1} & 12.8 & \textbf{15.9} & \textbf{24.1} & \textbf{20.1} & \textbf{22.9} & \textbf{19.9} \\
    LiveKit v1-mini & 24.4 & 24.8 & 27.8 & 26.0 & 22.4 & \textbf{27.4} & 43.5 & 33.7 & 23.5 & \textbf{12.3} & 32.9 & 36.9 & 35.3 & 27.5 & 28.4 \\
    ultraVAD & 45.5 & 35.5 & 27.7 & 41.6 & 34.4 & 39.1 & 45.0 & 40.4 & 12.0 & 33.8 & 42.3 & 42.8 & 44.9 & 38.8 & 37.4 \\
    Smart Turn v3.2 & 39.2 & 35.5 & 35.2 & 38.7 & 38.9 & 42.6 & 46.2 & 44.1 & 34.9 & 30.1 & 42.3 & 42.7 & 43.3 & 42.9 & 39.7 \\
    VAP & 54.6 & 53.6 & 47.0 & 52.6 & 54.5 & 49.5 & 55.3 & 50.3 & 48.8 & 43.5 & 42.3 & 55.3 & 50.4 & 51.2 & 50.6 \\
    \midrule
    \multicolumn{16}{@{}l}{\textit{Latency at 5\% cut-offs (ms)}} \\
    LiveKit v1 (closed) & 706 & 672 & 543 & 643 & 635 & 454 & 832 & 785 & 321 & 324 & 610 & 703 & 718 & 799 & 625 \\
    \ours \textbf{VoxTurn v1} & \textbf{751} & \textbf{801} & \textbf{669} & \textbf{782} & \textbf{662} & \textbf{606} & \textbf{913} & \textbf{912} & \textbf{433} & \textbf{542} & \textbf{728} & \textbf{796} & \textbf{898} & \textbf{814} & \textbf{736} \\
    LiveKit v1-mini & 811 & 947 & 1071 & 926 & 798 & 624 & 938 & 966 & 766 & 712 & 884 & 897 & 1080 & 919 & 881 \\
    ultraVAD & 807 & 864 & 899 & 986 & 846 & 616 & 939 & 1005 & 462 & 731 & 910 & 849 & 1075 & 926 & 851 \\
    Smart Turn v3.2 & 814 & 963 & 1051 & 980 & 873 & 639 & 945 & 980 & 837 & 734 & 884 & 836 & 1013 & 959 & 893 \\
    VAP & 839 & 1068 & 1131 & 1067 & 917 & 655 & 1009 & 1012 & 796 & 800 & 884 & 899 & 1120 & 942 & 938 \\
    \bottomrule
  \end{tabular}
\end{table}

\section{Training hyperparameters}
\label{app:hparams}

\begin{table}[!h]
  \caption{\vt{} v1 training configuration.}
  \label{tab:hparams}
  \centering\small
  \begin{tabular}{@{} l l @{}}
    \toprule
    \textbf{Setting} & \textbf{Value} \\
    \midrule
    Backbone & Qwen3-ASR-0.6B (AuT encoder frozen, Qwen3-0.6B language model) \\
    Trained parameters & projector, LoRA, linear head (11.8M) \\
    LoRA & rank 16, $\alpha = 32$, every linear layer of the language model \\
    Audio context & 8\,s, left-padded with silence; 13 tokens/s (104 tokens) \\
    Training samples & 300k \\
    Batch & 128 (4 micro-batches of 32) \\
    Optimiser & AdamW, lr $2\times10^{-4}$, $\beta=(0.9, 0.98)$, weight decay 0.01 \\
    Schedule & 100 warm-up steps, cosine decay to 10\% \\
    Gradient clipping & 1.0 \\
    Precision & bf16 autocast \\
    Loss weights & $\lambda_{\text{aux}} = \lambda_{\text{cons}} = 0.5$ \\
    Decision time in a pause & uniform over the first $\min(d, 1.5\,\text{s})$ \\
    Compute & 78 minutes on one NVIDIA L40S \\
    \bottomrule
  \end{tabular}
\end{table}

\section{TurnBench decision policy}
\label{app:policy}

For EOT we add a two-channel \emph{decision policy}: each silence opens a candidate, committed at the first frame
where
\begin{enumerate}[label=(\roman*)]
  \item the silence has lasted $d_{\min}$ and $p(\text{eot})\,\big(1-p(\text{resume} \le 0.5\,\text{s})\big) \ge
    \theta_{\text{own}}$, or
  \item the partner, starting at most 1\,s before or during the silence, has spoken for $m$\,s and \vt{} reads it as
    floor-taking ($p(\text{not backchannel})\,p(\text{not invalid}) \ge \theta_{\text{floor}}$), or
  \item a deadline passes.
\end{enumerate}
Events wait 0.5\,s after the speech offset. As a pause costs at most one false positive, a confirming event follows
at silence $t_{\text{conf}}$, and the last silence ends with an event at the end of the audio. The ten parameters are
random-searched for the best dev recall at FPR $\le 0.1$.

\end{document}